\documentclass[aps,prb,twocolumn,superscriptaddress,floats,showpacs]{revtex4}
\usepackage{}
\usepackage{txfonts}
\usepackage{amssymb}
\usepackage{graphicx}

\begin{document}

\title{The unique electronic structure of  Ca$_{10}$(Pt$_4$As$_8$)(Fe$_{2-x}$Pt$_x$As$_2$)$_5$ with metallic  Pt$_4$As$_8$  layers}

\author{X. P. Shen}

\author{S. D. Chen}

\author{Q. Q. Ge}

\author{Z. R. Ye}

\affiliation{State Key Laboratory of Surface Physics, Department of Physics,  and Advanced Materials Laboratory, Fudan University, Shanghai 200433, People's Republic of China}

\author{F. Chen}

\affiliation{Hefei National Laboratory for Physical Science at Microscale and Department of Physics, University of Science and Technology of China, Hefei, Anhui 230026, People's Republic of China}

\author{H. C. Xu}

\affiliation{State Key Laboratory of Surface Physics, Department of Physics,  and Advanced Materials Laboratory, Fudan University, Shanghai 200433, People's Republic of China}

\author{S. Y. Tan}

\author{X. H. Niu}

\author{Q. Fan}

\author{B. P. Xie}

\author{D. L. Feng}\email{dlfeng@fudan.edu.cn}

\affiliation{State Key Laboratory of Surface Physics, Department of Physics,  and Advanced Materials Laboratory, Fudan University, Shanghai 200433, People's Republic of China}

\date{\today}

\begin{abstract}
We studied the low-lying electronic structure of the newly discovered iron-platinum-arsenide superconductor, Ca$_{10}$(Pt$_4$As$_8$)(Fe$_{2-x}$Pt$_x$As$_2$)$_5$ ($T_c$ = 22~K) with angle-resolved photoemission spectroscopy. We found that the  Pt$_4$As$_8$ layer contributes to a small electron-like Fermi surface, indicative of metallic charge reservoir layers that are rare for iron based superconductors.  Moreover, the  electronic structure of the FeAs-layers resembles those of other prototype iron pnictides to a large extent. However, there is only $d_{xy}$-orbital originated hole-like Fermi surface near the zone center, which is rather unique for an iron pnictide superconductor with relatively high superconducting transition temperature; and  the $d_{xz}$ and $d_{yz}$ originated bands are not degenerate at the zone center.
Furthermore, all bands near the Fermi energy show negligible $k_z$ dependence, indicating the strong two-dimensional nature of this material.
Our results indicate this material possesses rather unique electronic structure, which enriches our current knowledge of iron based superconductors.
\end{abstract}


\maketitle

\section{introduction}

The discovery of iron based superconductors with high transition temperature ($T_c$; maximally $\sim$ 55~K \cite{55K1, 55K2}) has attracted considerable interest. Most iron pnictides  have quasi-two dimensional FeAs layers where superconductivity occurs. Extensive experimental and theoretical studies have demonstrated that the low-energy electronic structure of iron pnictides are dominated by multiple Fe-3d orbitals \cite{Fe3dorbital1, Fe3dorbital2, Fe3dorbital3}. The main difference among them lies in the structure of spacer layers between FeAs layers, which mostly contain alkaline metals, alkaline earth metals, rare-earth oxides or alkaline-earth fluorides \cite{charge resevoir}. These spacer layers serve as $``$charge reservoirs$"$ as in cuprates and are known to be insulating due to strong ionic bonds. It has been argued that higher $T_c$ of cuprate superconductors are facilitated by enhanced CuO$_2$ plane coupling through the (Bi,Tl,Hg)-O intermediary layers \cite{enhacedCu, Tl1, Tl2, HgO}. Then analogous to cuprate, one possible approach for realizing higher $T_c$ in the iron based superconductors is to explore novel spacer layers to tune the electronic states of the FeAs layers \cite{SrVOFeAs1, CaAlOFeAs}.

% Fig.1 sample picture & Laue pattern

\begin{figure}[t]
\includegraphics[width=8.6cm]{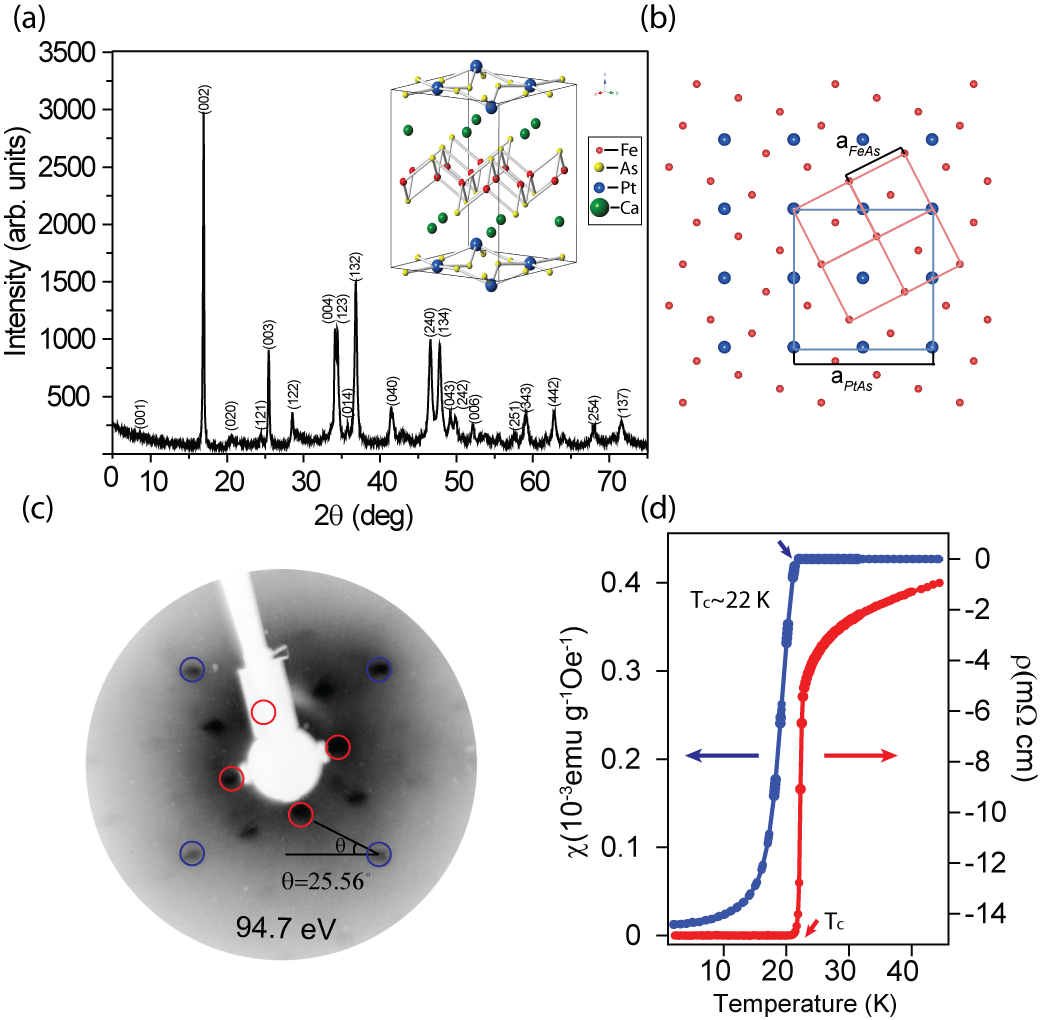}
\caption{The crystallographic and superconducting properties of Ca$_{10}$(Pt$_4$As$_8$)(Fe$_{2-x}$Pt$_x$As$_2$)$_5$. (a)  Powder XRD pattern of Ca$_{10}$(Pt$_4$As$_8$)(Fe$_2$As$_2$)$_5$. The powder was obtained by grinding the single crystals. The inset is a schematic picture of the crystal structure of Ca$_{10}$(Pt$_4$As$_8$)(Fe$_2$As$_2$)$_5$. (b) Top view of Pt$_4$As$_8$ and FeAs layers. (c) The Low-energy electron diffraction pattern of a Ca$_{10}$(Pt$_4$As$_8$)(Fe$_{2-x}$Pt$_x$As$_2$)$_5$ single crystal. Reflection peaks from both FeAs (blue circles) and Pt$_4$As$_8$ (red circles) lattices are observed. (d) The magnetic susceptibility of an as-grown Ca$_{10}$(Pt$_4$As$_8$)(Fe$_{2-x}$Pt$_x$As$_2$)$_5$ single crystal taken at a magnetic field of 20~Oe in the zero-field cool mode, and its resistivity as a function of temperature. }
\label{sample}
\end{figure}

% Fig.2 polarized map and band structure

\begin{figure*}[t]
\includegraphics[width=16cm]{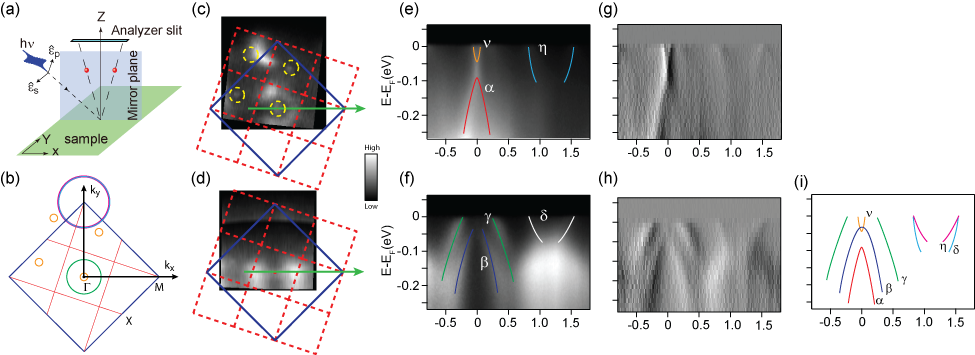}
\caption{Photoemission data of Ca$_{10}$(Pt$_4$As$_8$)(Fe$_{2-x}$Pt$_x$As$_2$)$_5$. (a) Schematic of two polarization geometries of experimental setups, reproduced from Ref. \cite{yanglexian}.  (b) Sketch of Fermi surface for Ca$_{10}$(Pt$_4$As$_8$)(Fe$_{2-x}$Pt$_x$As$_2$)$_5$.  (c) Photoemission intensity map at the Fermi energy ($E_F$) integrated over [$E_F$-10 meV, $E_F$+10 meV] in $p$ polarization. The 2D Brillouin zone in blue is for FeAs layer and the dash one in red is for PtAs layer.  (e) Photoemission intensity along (0, 0)--($\pi$, 0) direction, as indicated by the green arrow in panel a. The solid lines indicate the bands dispersions extracted from the corresponding MDC's. (g) The second derivative with respect to energy of the data in panel a.  (d), (f), and (h) are the same as in panels c, e, and g, respectively, but taken in $s$ polarization.  (i) Summary of the experimental band structure for Ca$_{10}$(Pt$_4$As$_8$)(Fe$_{2-x}$Pt$_x$As$_2$)$_5$.  All data were taken with 26~eV photons at UVSOR. }
\label{pomap}
\end{figure*}

Recently, superconductivity up to 38~K was reported in the quaternary Ca-Fe-Pt-As system \cite{PNAS 38K, 38K}. This system has FeAs layers similar to other iron based superconductors but the  spacer layers  are composed of two Ca layers and one PtAs layer sandwiched between them. As for PtAs layers, two structures have been identified, Pt$_3$As$_8$ and Pt$_4$As$_8$. The two corresponding systems are Ca$_{10}$(Pt$_3$As$_8$)(Fe$_{2-x}$Pt$_x$As$_2$)$_5$ (named as the 10-3-8 compound) and Ca$_{10}$(Pt$_4$As$_8$)(Fe$_{2-x}$Pt$_x$As$_2$)$_5$ (named as the 10-4-8 compound), with maximal $T_c \sim$15~K \cite{bandcalculation2} and 38~K \cite{PNAS 38K, 38K}, respectively. One of the most intriguing features of these materials is that both Pt$_n$As$_8$ (n=3, 4) and Fe$_2$As$_2$ blocks would compete for the electrons provided by the Ca atoms \cite{negative valence}. Based on Zintl$'$s chemical concept, Ca$_{10}$(Pt$_3$As$_8$)(Fe$_2$As$_2$)$_5$ is a valence satisfied compound with the charges of the [Pt$_3$As$_8$]$^{10-}$ layer perfectly balanced by the [Ca$_{10}$]$^{20+}$ and [Fe$_{10}$As$_{10}$]$^{10-}$ layers, and thus the Pt$_3$As$_8$ layer is expected to be semiconducting \cite{valence satisfied compound, PNAS 38K}. On the other hand, because of one more Pt atom in the PtAs layer, the [Pt$_4$As$_8$]$^{12-}$ layer in Ca$_{10}$(Pt$_4$As$_8$)(Fe$_2$As$_2$)$_5$ is expected to be metallic \cite{PNAS 38K}. The metallic Pt$_4$As$_8$ layers would likely enhance the interlayer coupling and might be responsible for the higher $T_c$  as compared to Ca$_{10}$(Pt$_3$As$_8$)(Fe$_{2-x}$Pt$_x$As$_2$)$_5$ \cite{shein}.
Theoretically, it is proposed that  the density of states (DOS) near the Fermi energy ($E_F$) for both phases has small contributions from the Pt states \cite{bandcalculation1,bandcalculation2}.  However,  a recent angle-resolved photoemission spectroscopy (ARPES) study on Ca$_{10}$(Pt$_3$As$_8$)(Fe$_{2-x}$Pt$_x$As$_2$)$_5$ ($T_c$ = 8~K) revealed no sign of Fermi pockets from the Pt$_3$As$_8$ layer and suggested that the Pt$_3$As$_8$  layer couples weakly to the FeAs layer \cite{ARPES1038}.
Another ARPES study on a 10-3-8 compound  ($T_c$ = 15~K), and a 10-4-8 compound  ($T_c$ = 35~K) also does not observe any Fermi pocket from the Pt$_3$As$_8$ or  Pt$_4$As$_8$ layers \cite{Bori2013}.

In this article, we present the  electronic structure of Ca$_{10}$(Pt$_4$As$_8$)(Fe$_{2-x}$Pt$_x$As$_2$)$_5$ ($T_c$ = 22~K), which is an electron overdoped 10-4-8 compound.
Our ARPES results reveal an electron Fermi pocket  around the zone center of the Pt$_4$As$_8$ Brillouin zone (BZ), which makes it the first iron based superconductor with metallic spacer layers.
Moreover, all the other bands resembles those of other iron pnictides, and their polarization dependencies indicate that their   orbital characters  are  similar to the corresponding bands in other iron pnictides as well.
Furthermore, we have studied the evolution of the band structure along $k_z$ direction in the three-dimensional (3D) momentum space. All bands show negligible $k_z$ dispersion, indicating the pronounced two-dimensionality of this material. Intriguingly, we found that there is only a $d_{xy}$-originated hole pocket around the  $\Gamma$-Z or (0,0,0)-(0,0,$\pi$) line, which is unique for a iron pnictide superconductor with such a high $T_c$.
The interactions between the Pt$_4$As$_8$  layer and the FeAs layer are manifested in the fact that  the $d_{xz}$ and $d_{yz}$ originated bands in the FeAs layers are no longer degenerate at $\Gamma$.  However,  we do not observe any  folding of the bands in one layer from the umklapp scattering of the crystal field of the other layer, which suggest the interaction is not strong. Our experimental results establish another member of the iron based superconductor family with an unique electronic structure, which enriches our current understanding of these materials.
Our results indicate that the charge reservoir layer could alter the electronic structure of iron based superconductors, and the superconducting properties might be enhanced further with novel metallic intermediary layers.

\section{sample properties and experimental setup}

Single crystals of Ca$_{10}$(Pt$_4$As$_8$)(Fe$_{2-x}$Pt$_x$As$_2$)$_5$ were synthesized with CaAs, FeAs, Pt powders as starting materials. The mixture with a ratio of CaAs:FeAs:Pt=1:1:0.5 was pressed into pellets, loaded into an alumina tube, and then sealed into an argon-filled iron crucible. The entire assembly was heated to 1423~K and kept for 72~h and then slowly cooled down to 1173~K at a rate of 2.5~K/h before shutting off the power. Energy dispersive X-ray spectroscopy (EDX) measurements give a chemical composition of
Ca : Fe : Pt : As= 2 : 1.905 : 0.912 : 3.586 for an as-grown sample. The ratio between Pt and As is similar to the previous report for another 10-4-8 compound \cite{PNAS 38K}, but with a higher Pt doping. The X-ray powder diffraction data are shown in Fig.~\ref{sample}(a), which give a=b=8.731~$\AA$, and c=10.537~$\AA$, similar to those reported in Ref.~\onlinecite{PNAS 38K}  as well.

The crystal structure of Ca$_{10}$(Pt$_4$As$_8$)(Fe$_2$As$_2$)$_5$ is shown in the inset of Fig.~\ref{sample}(a), consisting of alternately stacked FeAs and Ca-Pt$_4$As$_8$-Ca layers \cite{PNAS 38K}. In contrast to the ``111", ``122" prototype iron pnictides, it has an additional Pt$_4$As$_8$ layer between the two Ca layers,  and the distance between the neighboring FeAs layer is among the largest in iron pnictides. Note that, the Ca atoms and the FeAs layers are arranged in the similar tetragonal structure to the ``111" iron pnictides, while the unique Pt$_4$As$_8$ layer matches the FeAs lattice constant with a$_{PtAs}$=$\sqrt{5}$ a$_{FeAs}$ and a rotation of $\theta$ = 25.56$^\circ$ [Fig.~\ref{sample}(b)], which is clearly shown by low energy electron diffraction (LEED) pattern in Fig.~\ref{sample}(c). Consequently, the unit cell of  Ca$_{10}$(Pt$_4$As$_8$)(Fe$_2$As$_2$)$_5$ is tetragonal.

Magnetic susceptibility measurements were performed with Quantum Design superconducting quantum interference device (SQUID) and in-plane resistivity measurements were performed with a Quantum Design physical property measurements system (PPMS). With Pt partially substituted for Fe in the FeAs layer in our as-grown Ca$_{10}$(Pt$_4$As$_8$)(Fe$_{2-x}$Pt$_x$As$_2$)$_5$ sample, resistivity and magnetic susceptibility measurements show clear signature of a superconducting transition at $T_c$ = 22~K [Fig.~\ref{sample}(d)]. The narrow transition temperature width  demonstrates the high quality of the Ca$_{10}$(Pt$_4$As$_8$)(Fe$_{2-x}$Pt$_x$As$_2$)$_5$ crystals.

% Fig.3 randomly polarized map to small electron pocket
\begin{figure*}[t]
\includegraphics[width=17cm]{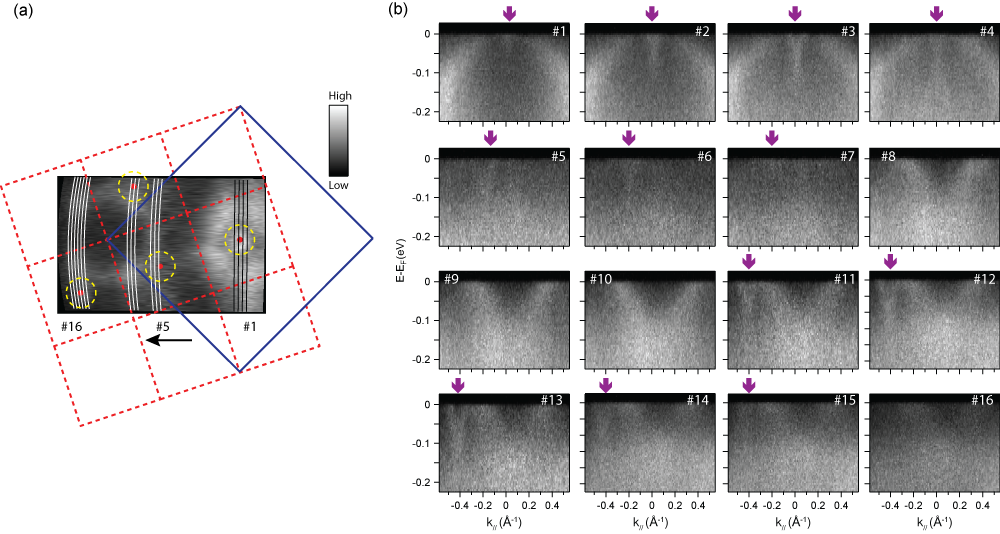}
\caption{ Photoemission data of Ca$_{10}$(Pt$_4$As$_8$)(Fe$_{2-x}$Pt$_x$As$_2$)$_5$. (a) Photoemission intensity map at $E_F$ integrated over [$E_F$-30 meV, $E_F$+30 meV]. The small electron pockets are enclosed by the dash circles.  (b) Photoemission intensity plots of cuts \#1-\#16 as indicated in panel a. The positions of the small electron pockets are marked by the arrows. All data were taken with randomly polarized 21.2~eV photons from a helium discharge lamp.}
\label{map}
\end{figure*}

ARPES measurements were performed at (1) Beamline 7U of the UVSOR synchrotron facility, with a variable photon energy and MBS A-1 electron analyzer, (2) the surface and interface spectroscopy Beamline of the Swiss Light Source (SLS), and (3) at an in-house system equipped  with  an  SPECS UVLS helium discharging lamp. Scienta electron analyzers are equipped in the (2) and (3) setups. The overall energy resolution was set to 15~meV or better, and the typical angular resolution was 0.3$^\circ$. The samples were pre-oriented by Laue diffraction, cleaved \emph{in situ} and then measured under ultrahigh vacuum better than 6$\times$10$^{-11}$ mbar. The experimental setup for polarization-dependent ARPES is shown in Fig.~\ref{pomap}(a). The incident
beam and the sample surface normal define a mirror plane.
For the the $s$ (or $p$) experimental geometries, the electric field of the incident photons is out of (or in) the mirror plane. The matrix element for the photoemission process could be described as

\begin{center}
$\left| M_{f,i}^\textbf{\emph{k}} \right|^2 \propto \left| \left\langle \Psi_f^\emph{\textbf{k}}\left| \hat{\varepsilon}  \cdot \emph{\textbf{r }} \right|\Psi_i^\emph{\textbf{k}} \right\rangle \right|^2$
\end{center}

\noindent Since the final state $\Psi _f^\emph{\textbf{k}}$ of photoelectrons could be approximated by a plane wave with its wave vector in the mirror plane,  $\Psi _f^\emph{\textbf{k}}$ is always even respect to the mirror plane in our experimental geometry. In the $s$ (or $p$) geometry, $\hat{\varepsilon}  \cdot \emph{\textbf{r }}$ is odd (or even) with respect to the mirror plane. Thus considering the spatial symmetry of the $3d$ orbitals, when the analyzer slit is along the high-symmetry directions, the photoemission intensity of specific even (or odd) component of a band is only detectable with the $p$ (or $s$) polarized light. For example, with respect to the mirror plane (the $xz$ plane), the even orbitals ($d_{xz}$, $d_{z^2}$ , and $d_{x^2-y^2}$ ) and the odd orbitals ($d_{xy}$ and $d_{yz}$) could be only observed in the $p$ and $s$ geometries, respectively  \cite{yanzhangBaCo, yanzhangNaFeAs}.

\section{Experimental results and analyses}

\begin{figure*}[t]
\includegraphics[width=17cm]{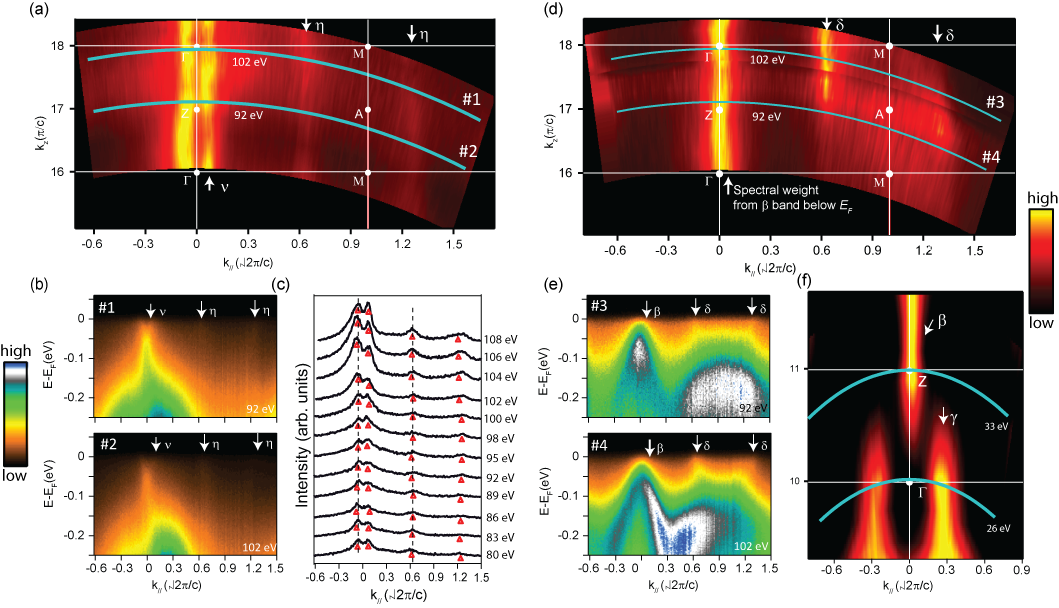}
\caption{The Fermi surface and band structure as a function of $k_z$ for Ca$_{10}$(Pt$_4$As$_8$)(Fe$_{2-x}$Pt$_x$As$_2$)$_5$.  (a) Photoemission intensity map in the $k_x$-$k_z$ plane in $p$ polarization. (b) The photoemission intensity with typical photon energies along the cuts \#1 and \#2 in panel a. (c) The corresponding MDC's at $E_F$ with different photon energies in panel a, the peaks indicating Fermi crossing $k_F$ are marked by red solid triangles. (d)-(e) are the same as in panels a-b, respectively, but in $s$ polarization. (f) is the same as in panel d, but taken at low photon energies from 22~eV to 38~eV for a better momentum resolution. Different $k_z'$s were accessed by varying the photon energies at Beamline of SLS and Beamline 7U of UVSOR, as indicated by the blue lines, where and inner potential of 13~eV is used to obtain $k_z$. The data in panels (a-e) and panel (f) were collected at SLS and UVSOR, respectively.}

\label{centerKz}
\end{figure*}

The photoemission intensity maps in the $p$ and $s$ polarizations are shown in Figs.~\ref{pomap}(c) and \ref{pomap}(d), respectively. As summarized in Fig.~\ref{pomap}(b), the observed Fermi surface consists of one hole pocket and one electron pocket around the zone center, two electron pockets around the zone corner of the FeAs BZ,  as well as several electron pockets located around the zone centers of extended Pt$_4$As$_8$ BZs. For simplicity, we will use the BZ according to the tetragonal FeAs lattice hereafter, if not specified. The photoemission intensity along (0, 0)--($\pi$, 0) under the $p$ and $s$ geometries are further displayed in Figs.~\ref{pomap}(e) and \ref{pomap}(f), respectively. Around the zone center, an electron band assigned as $\nu$ together with a fast dispersive band $\alpha$ below $\sim$~90~meV is resolved in the $p$ polarization. The $\nu$ band crosses $E_F$, forming a small electron pocket. On the other hand, two bands ($\beta$ and $\gamma$) show up in the $s$ polarization. The $\gamma$ band crosses $E_F$, forming a hole pocket around the zone center, while the band top of $\beta$ is about 35~meV below $E_F$. With the assistance from the second derivative of the photoemission intensity with respect to energy, two electron bands around the zone corner, $\eta$ and $\delta$, are clearly identified in $p$ and $s$ geometries respectively, exhibiting opposite spatial symmetries.

In general, as recapitulated in Fig.~\ref{pomap}(i), most of the bands resemble the band structures in the prototype iron pnictides except the electron-like band ($\nu$) around the zone center, which demonstrates that the low-lying electronic structures of Ca$_{10}$(Pt$_4$As$_8$)(Fe$_{2-x}$Pt$_x$As$_2$)$_5$ are mainly from the Fe-$3d$ orbitals. Moreover, the polarization dependencies for these bands are the same as those in NaFeAs and other iron pnictides. Therefore, according to previous knowledge on iron pnictides \cite{yanzhangBaCo, yanzhangNaFeAs}, we can ascribe their orbital characters along the $\Gamma-M$ direction as follow: \begin{enumerate}
 \item  $\alpha$  around the zone center could be ascribed to the even $d_{xz}$ orbital;
 \item   $\beta$   could be ascribed to the odd $d_{yz}$ orbital;
 \item  $\gamma$   could be ascribed to the $d_{xy}$ orbital;
 \item $\delta$ around the zone corner  could be ascribed to the $d_{xy}$ orbital;
 \item $\eta$  could be ascribed to the $d_{xz}$ orbitals.
 \end{enumerate}

% Fig.3 randomly polarized map to small electron pocket

Despite the overall similarity, there are still obvious differences between the Fermi surface sheets of Ca$_{10}$(Pt$_4$As$_8$)(Fe$_{2-x}$Pt$_x$As$_2$)$_5$ and those of the prototype iron pnictides.
First of all,  the band tops of the $d_{xz}$ dominated $\alpha$ band and the $d_{yz}$ dominated $\beta$ band are not degenerate at $\Gamma$ (Fig.~\ref{pomap}(i)). This means that the rotational symmetry of the tetragonal FeAs layer is broken, which might be due to the crystalline potential imposed by the Pt$_4$As$_8$ layer. Secondly,  besides the small $\nu$ electron pocket around $\Gamma$, several bright features are further observed, as enclosed by dash lines in Fig.~\ref{pomap}(a). These features are more obvious in the data taken with randomly polarized 21.2 eV photons  in Fig.~\ref{map}, where four small electron pockets are observed and they match well with the BZs of Pt$_4$As$_8$ layer.
The momentum evolution of the electron pockets are revealed by the photoemission intensity plots in Fig.~\ref{map}(b), particularly along the cuts \#1-\#4 and \#11-\#16.
The dispersions of these electron-like bands are identical as expected. Based on these observations, the $\nu$ Fermi pocket should be attributed to the states in the Pt$_4$As$_8$ layer. This finding  qualitatively agrees with  a recent first principles band calculation \cite{bandcalculation1}, which suggests that the Pt$_4$As$_8$ layers contribute to small electron-like bands at the zone center.

% Fig.4 Kz map near zone center

To understand the detailed electronic structure evolution along $k_z$ direction in 3D momentum space, the photon energy dependent ARPES measurements were conducted. The measured Fermi surface cross-section in the $k_x$-$k_z$ plane for both $p$ and $s$ polarizations are shown in Figs.~\ref{centerKz}(a) and \ref{centerKz}(d), respectively. The cross-sections of the $\nu$, $\eta$ and $\delta$ Fermi surfaces are nearly straight cylinders along $k_z$ direction, thus indicating weak $k_z$ dependence. Note that in Fig.~\ref{centerKz}(d), the strong spectral weight around zone center originates from the band top of $\beta$ below $E_F$ and the $\gamma$ band is hardly visible, likely due to matrix element effects. As shown in Fig.~\ref{centerKz}(f), the $\gamma$ band is better revealed at lower photon energies and negligible $k_z$ dispersion is observed. Therefore, all bands show negligible $k_z$ dispersion, indicating the strong two-dimensional (2D) nature of the electronic states. The two dimensional nature is an expected  consequence of the thick Ca-Pt$_4$As$_8$-Ca  spacer layer between the neighboring two FeAs layers, and similar experimental results were observed in LaOFeAs \cite{yanglexian}. Moreover, we note that there is an ellipsoidal electron Fermi surface around Z in  heavily electron doped  iron pnictides such as LiFe$_x$Co$_{1-x}$As or NaFe$_x$Co$_{1-x}$As \cite{zirongyeNaCo}, the negligible $k_z$ dispersion of $\nu$ here again shows that it is from the  Pt$_4$As$_8$ layer.

\section{discussions and conclusions}

Based on the above analysis, the most unique electronic structure of the 10-4-8 compound studied here is its
electron pocket around zone center of the Pt$_4$As$_8$ BZ, which makes it the first experimentally proven iron pnictide whose  spacer layer is metallic and likely participates in superconductivity.
The recent ARPES measurements of another 10-4-8 compound ($T_c$=35 K) did not find such an electron pocket \cite{Bori2013}. Based on the Fermi surface volume, it is clear that our sample is doped with more electrons. Moreover, since  Fermi surface originated from the Pt$_3$As$_8$ layer has never been observed
for the 10-3-8 compounds with much lower $T_c$ \cite{ARPES1038,Bori2013}, the metallicity of the spacer layer might not be so relevant to the superconductivity in these Ca-Fe-Pt-As compounds.
On the other hand, due to proximity effect, the superconductivity in the FeAs layer will also result in superconductivity in the Pt$_4$As$_8$ layer. The pairing behaviors on the $\nu$ pocket deserves further investigation, and  we leave this for the future  studies.

When considering the impact of Pt$_4$As$_8$ layer on FeAs layer, we notice that the $d_{xz}$/$d_{yz}$ orbital is no longer equivalent and are not degenerate at $\Gamma$. This suggests that the crystalline potential arising from Pt$_4$As$_8$ layer  breaks the rotational symmetry of FeAs tetragonal lattice. On the other hand, the crystalline potential of the Pt$_4$As$_8$ or FeAs layer does not cause any observable umklapp scattering, which would result in band folding \cite{hongwei}. However, we notice that such a non-degenerate behavior of  the $d_{xz}$/$d_{yz}$ bands  was observed in a 10-3-8 compound but not in a 10-4-8 compound \cite{Bori2013}. This disagreement might be due to some sample-specific issues that needs further exploration.

The electronic structure  of the FeAs-layer in this 10-4-8 compound ($T_c$=22 K) is somewhat similar to that of  LiFeAs \cite{zirongyeNaCo}, which has only one large hole-like Fermi surface pockets made of the $d_{xy}$ orbital around $\Gamma$.  However for LiFeAs, there is a hole Fermi pocket made of $d_{xz}/d_{yz}$ orbitals around Z, which was considered to be crucial for the superconductivity. It was found that the superconductivity disappears following the disappearance of the $d_{xz}/d_{yz}$-based small hole-like Fermi pocket  with increased cobalt doping. Meanwhile, the  $d_{xy}$-based band was found to be heavily scattered by the cobalt impurity \cite{zirongyeNaCo}, which might explain why it could not sustain the superconductivity.  Similar behaviors have been found in  BaFe$_{2-x}$Co$_x$As$_2$ as well \cite{Kaminski}, where the $d_{xy}$-based hole Fermi surface was considered not to be able to sustain the superconductivity.
In contrast, although there is only  $d_{xy}$-based hole like Fermi surface in the  10-4-8 compound studied here, its $T_c$ is still relatively high, which is in coherence with its relatively small scattering rate. Such an iron pnictide superconductors with only  a $d_{xy}$-based hole Fermi surface at the zone center is rather unique.

To summarize, we have reported the electronic structure of Ca$_{10}$(Pt$_4$As$_8$)(Fe$_{2-x}$Pt$_x$As$_2$)$_5$ by ARPES and resolved a rather unique Fermi surface topology.
Its electronic structure is rather   two-dimensional due to its  unusually thick spacer layer.
We have observed the Fermi surface of the Pt$_4$As$_8$ layer, which makes it the first iron based superconductor with  metallic spacer layers.  The  Fermi surfaces of the FeAs layers  show good correspondence to those of other prototype iron pnictides, except there is only a $d_{xy}$-based hole Fermi cylinder. Our data thus establish an unique  iron based superconductor with a particular electronic structure, which deepens and enriches the current understanding of these materials.

\section{acknowledgements}

Part of this work was performed at the Surface and Interface Spectroscopy Beamline, SLS, Paul Scherrer Institute, Villigen, Switzerland and Beamline 7U of the UVSOR synchrotron facility, Institute for Molecular Science and The Graduate University for Advanced Studies, Okazaki, Japan. This work is supported by the  National Basic Research Program of China  (973 Program) under the grant Nos. 2012CB921402, 2011CB921802, 2011CBA00112, and National Science Foundation of China.

\end{document}